\documentclass[manuscript, nonacm=true]{acmart}
\makeatletter
\def\@ACM@checkaffil{
    \if@ACM@instpresent\else
    \ClassWarningNoLine{\@classname}{No institution present for an affiliation}%
    \fi
    \if@ACM@citypresent\else
    \ClassWarningNoLine{\@classname}{No city present for an affiliation}%
    \fi
    \if@ACM@countrypresent\else
        \ClassWarningNoLine{\@classname}{No country present for an affiliation}%
    \fi
}
\makeatother
\usepackage{array}
\usepackage{booktabs}
\usepackage{tabularx}
\usepackage{multirow}
\usepackage{wrapfig}

\usepackage{gensymb}

\usepackage{amsmath}
\usepackage{amsthm}
\usepackage{algorithm2e}
\usepackage{algorithmic}
\usepackage{bigints}

\usepackage{epsfig}
\usepackage{graphicx}
\usepackage{subcaption}

\usepackage[normalem]{ulem}
\usepackage{ifthen}
\usepackage{enumerate}
\usepackage{enumitem}
\usepackage{listings}
\usepackage{textcomp}
\usepackage{xcolor}
\usepackage{titlesec}
\usepackage{url}

\usepackage{pdfpages}

\definecolor{dkred}{rgb}{0.7,0,0}
\definecolor{dkpurple}{HTML}{4e02eb}
\definecolor{dkgreen}{HTML}{006329}
\definecolor{teal}{HTML}{006482}
\definecolor{string}{HTML}{02782d}
\definecolor{Fuchsia}{HTML}{8C368C}
\definecolor{bar1}{HTML}{00cc00}
\definecolor{bar2}{HTML}{009900}
\definecolor{bar3}{HTML}{00cc60}
\definecolor{bar4}{HTML}{009960}
\definecolor{bar5}{HTML}{60cc00}
\definecolor{bar6}{HTML}{609900}

\newif\ifdraft\drafttrue

\newcommand{\pages}[1]{}

\usepackage{ifthen}
\usepackage{pgfgantt}

\newwrite\workplanfile
\immediate\openout\workplanfile=\jobname.workplan

\gdef\currentganttsection{}

\ganttset{bar height=.6}
\usepackage{anyfontsize}

\newcommand{\timelinechart}{
   \immediate\closeout\workplanfile
   \begin{ganttchart}[
        expand chart=\textwidth,
        y unit chart=.4cm,
      ]{1}{7}
    \gantttitle{Year 1}{1}
    \gantttitle{Year 2}{1}
    \gantttitle{Year 3}{1}
    \gantttitle{Year 4}{1}
    \gantttitle{Year 5}{1}
    \gantttitle{Year 6}{1}
    \gantttitle{Year 7}{1}
    \renewcommand{\tasksection}{\tasksectiongantt}
    \input{\jobname.workplan}
  \end{ganttchart}
}

\gdef\timelinecolor{lightgray}

\newcommand{\tasksectiongantt}[5]{%
  \ganttnewline
  \ganttbar
       [bar/.append
          style={fill=\timelinecolor}]
       {\hbox to 3in{{\footnotesize {\bf \currentganttsection}
                     \hfill
                     #1}}}
       {#2}{#3}
  \gdef\currentganttsection{}
}
\newcommand{\tasksection}[5]{\tasksectiongantt{#1}{#2}{#3}{#4}{#5}}

\newcommand{\timelinedetails}{
  \immediate\closeout\workplanfile
  \begin{tabular}[c]{@{}lll}
    {\bf Themes and Tasks} & {\bf Main effort} & {\bf Supporting effort}
    \\
    \hline
    \input{\jobname.workplan}
  \end{tabular}
}

\newcommand{\tasksectionroles}[5]{%
  \ifthenelse{\equal{\currentganttsection}{}}{}{{\bf \currentganttsection}\\}%
  \gdef\currentganttsection{}%
  \hspace{2em} #1 & #4 & #5 \\
}

\newcommand{\proposalName}{ 
Carbon Connect: An Ecosystem for Sustainable Computing
}
\title{\proposalName}

\author{Benjamin C. Lee}
\affiliation{University of Pennsylvania}

\author{David Brooks}
\affiliation{Harvard University}

\author{Arthur van Benthem}
\affiliation{University of Pennsylvania}

\author{Udit Gupta}
\affiliation{Cornell University}

\author{Gage Hills}
\affiliation{Harvard University}

\author{Vincent Liu}
\affiliation{University of Pennsylvania}

\author{Linh Thi Xuan Phan}
\affiliation{University of Pennsylvania}

\author{Benjamin Pierce}
\affiliation{University of Pennsylvania}

\author{Christopher Stewart}
\affiliation{Ohio State University}

\author{Emma Strubell}
\affiliation{Carnegie Mellon University}

\author{Gu-Yeon Wei}
\affiliation{Harvard University}

\author{Adam Wierman}
\affiliation{California Institute of Technology}

\author{Yuan Yao}
\affiliation{Yale University}

\author{Minlan Yu}
\affiliation{Harvard University}

\authorsaddresses{\textbf{Corresponding Authors.} Benjamin C. Lee, University of Pennsylvania, leebcc@seas.upenn.edu; David Brooks, Harvard University, dbrooks@g.harvard.edu.}

\thanks{\textbf{Acknowledgements.} U.S. National Science Foundation Expeditions in Computing: Carbon Connect -- An Ecosystem for Sustainable Computing.}

\renewcommand{\shortauthors}{Lee, Brooks, et al.}

\begin{document}

\begin{abstract}
\end{abstract}

\maketitle
\makeatletter \gdef\@ACM@checkaffil{} \makeatother

Information and communication technology (ICT) accounts for a shockingly large share of global greenhouse gas (GHG) emissions---estimates range from 2.1\% to 3.9\%~\cite{FREITAG2021100340}. To address this grand challenge, the International Telecommunication Union targets a 45\% reduction in ICT emissions by 2030~\cite{itu_carbon}, in line with the Paris Agreement’s goal to limit warming to 1.5°C above pre-industrial levels. Satisfying demands for computing while meeting these goals will be difficult and expensive, demanding rigorous methods and solutions that balance sustainability benefits against implementation costs. To succeed, computer scientists, electrical engineers, environmental scientists, and economists must develop an {\it ecosystem for sustainable computing} with rigorous, transformative solutions to computing's carbon problem, responding to the powerful call for action from Knowles et al.~\cite{10.1145/3503916}: computing must end the ``digital exceptionalism'' that brushes aside its own carbon footprint because of the productivity and efficiency it provides to society. 

We envision four broad research thrusts that are needed to produce design and management strategies for sustainable next-generation computer systems. First, we require accurate models for carbon accounting and reporting in computing technology. Second, for the embodied carbon emitted during the manufacture of hardware and infrastructure, we must adopt life-cycle design strategies that more effectively reduce, reuse and recycle hardware at scale. Third, for operational carbon associated with computing's electricity use, we must not only embrace renewable energy but also manage systems to use that energy more efficiently. And fourth, we require integrated, cross-cutting strategies for hardware design and management because techniques that reduce operational carbon may increase embodied carbon and vice versa.

New hardware design and management strategies must be developed in context. These strategies must seek to flatten and then reverse growth trajectories for computing power and carbon for society's most rapidly growing applications such as artificial intelligence. These strategies must also be cognizant of economic policy and regulatory landscape, aligning private initiatives with societal goals. Many of these broader goals will require computer scientists to develop deep, enduring collaborations with researchers in economics, law, and industrial ecology to spark changes in broader practice.

\section{Driving Applications}

Advances in artificial intelligence (AI) are enabled by massively scaling deep models and their training data \cite{ghorbani2022scaling,wei2022emergent}, which in turn impact sustainability \cite{strubell-etal-2019-energy, dodge22measuring}. Benchmarking AI's carbon footprint for model development, training, and deployment could help researchers identify the most pressing challenges. An integrated hardware-software perspective will be particularly helpful as AI is in the midst of a hardware lottery: dominant models may be those that benefit most from hardware trends \cite{hooker21}. Researchers should explore the net impact of custom hardware, which could reduce operational carbon through energy efficiency but increase embodied carbon through semiconductor manufacturing. 

The future of sustainable AI hinges on its ability to adapt in response to the varying availability of resources such as data, hardware, and electricity. First, there is a compelling need to design, train, and deploy AI models that offer performance, efficiency, and accuracy on a broad spectrum of hardware platforms. Such models would ensure backward compatibility for and equity of access to AI features, permitting users to slow the rate of hardware refreshes due to sustainability or financial constraints. Instead of deprecating systems after just a few years, how can we develop models and platforms that remain relevant over longer periods and better amortize the carbon costs of model training and deployment?

Second, there is a complementary need for programmable, reconfigurable hardware that support a broad spectrum of AI workloads. Such processors would allocate precisely the hardware required for data processing, training, or inference, consuming energy in proportion to utilization. Instead of designing static AI accelerators, how can we develop flexible, general processors that are relevant for large classes of AI computation and better amortize embodied carbon from semiconductor fabrication? Finally, demand response strategies would allow models and platforms to modulate their use of electricity based on its carbon intensity. To what extent can AI workloads be scheduled across time and space to reduce operational carbon?

If successful, this research agenda will reverse current trends and permit advanced AI with lower carbon costs. Google consumes 1.5-2.3TWh for AI, 10--15\% of its total energy use~\cite{patterson22}. Meta attributes 30\% of its AI energy for data processing, 30\% to model training, and 40\% to inference~\cite{wu22}. Studies for BLOOM's 176B-parameter language model, a GPT-3 replica, are alarming. Training uses 433MWh and emits 25T-CO$_2$e whereas inference uses 914KWh and emits 19kgs-CO$_2$e per day assuming 558 requests per hour \cite{luccioni2022estimating}. Production models' carbon footprints could easily be 1000x higher assuming one query from each of ChatGPT's 13M daily unique visitors in January 2023 and considering interest in applications of this technology.

\section{Carbon Accounting}

Computing's embodied carbon is incurred during hardware manufacturing. Modeling embodied carbon is exceptionally difficult because already complex semiconductor fabrication processes are evolving to accommodate emerging technologies such as nanomaterials \cite{hills18a, sabry_aly18}, photonic devices \cite{wang2018integrated}, and advanced heterogeneous integration \cite{lau22_advanced_packaging, sabry_aly18}. Yet we are optimistic because the manufacture of ``new'' technologies actually leverage many existing process flows. By mixing and matching steps in mature flows---lithography, metal and oxide deposition, etching, thermal annealing, \textit{etc.}---we might estimate carbon for flows not yet in production. For example, the first monolithic 3D process flow that integrates next-generation transistors and RRAM was recently deployed in a commercial foundry, SkyWater \cite{srimani_and_hills20}. This "new" flow re-orders existing steps and adds one new step for depositing 1D semiconductors.

Operational carbon depends on the amount and carbon intensity of the energy consumed. We must design energy profilers for individual tasks, helping operators track energy usage and guide management. System telemetry will be combined with grid telemetry, which details renewable energy generation across time. But estimating electricity's carbon intesnity is non-trivial. The marginal emission rate, which depends on the most recently activated generation source on the grid, may overstate operational carbon because datacenters often negotiate purchase agreements and receive credits from their investments in renewable energy and because grid operators often transfer energy across boundaries of regional balancing authorities. 

Telemetry lays foundations for attribution, which assigns responsibility for carbon to individual pieces of computation. A task's operational carbon depends on fine-grained energy telemetry and allocation of shared datacenter overheads. Estimating a task's share of embodied carbon for servers and datacenter infrastructure requires sophisticated analysis. Servers co-locate tasks and each task may use heterogeneous mixes of hardware that impact its share of embodied carbon. Game theory and the Shapley value may provide frameworks for fair attribution \cite{llull17}. 

We require reliable, harmonized, and transparent carbon accounting methods. Energy use and its emissions are verifiable through the EPA's carbon intensity statistics for power plants and directly measured energy for hardware components. Estimating energy use for fabrication is more difficult but can leverage published sustainability reports and datasets. Carbon from chemical and fuel combustion during fabrication benefits from a good understanding of the chemistry. Carbon accounting often leverages life cycle assessment (LCA) methods. Open-source models would lay the foundations for improving analysis and engaging stakeholders \cite{sleep21}. Such efforts are far behind in computing but other industries have harmonized accounting. For example, the EPA and California set standards to reduce GHG emissions from fuels, using open-source tools \cite{GREET, ElHoujeiri2013OpensourceLT}. 

\textbf{Risks.} Accounting frameworks and models are only as good as the data they ingest. Data for embodied carbon could be derived from first principles and industrial datasets. Tools, such as ACT and imec.netzero, estimate yields from integrated circuit (IC) manufacturing, where industry data is closely guarded, using broad and parameterized models (\textit{e.g.}, Murphy, Poisson). Moreover, techniques in robust optimization can account for uncertainty from input data and models to produce feasible solutions with accompanying confidence intervals.

Validating embodied carbon models also present challenges. But we draw inspiration from the IC community, which has already developed a combined bottom-up, top-down approach for performance and power. Models at various levels -- from simulating transistor physics to modeling datacenter power -- shape our overall understanding of the industry. Similarly, we envision a standard approach to modeling embodied carbon. At the bottom, we could model energy used by individual fabrication process steps and machines. At the middle, we could model energy used and emissions produced by the fabrication facility. At the top, we could estimate overall production volumes and carbon footprints with published sustainability reports from industry leaders. As with IC models, we expect validation and accuracy to improve over time as more data become available and models are refined.

\section{Embodied Carbon}

\begin{figure}
\begin{minipage}[t]{0.48\linewidth}
\centering
    \includegraphics[width=\textwidth, trim = 0mm 0mm 0mm 0mm, clip]{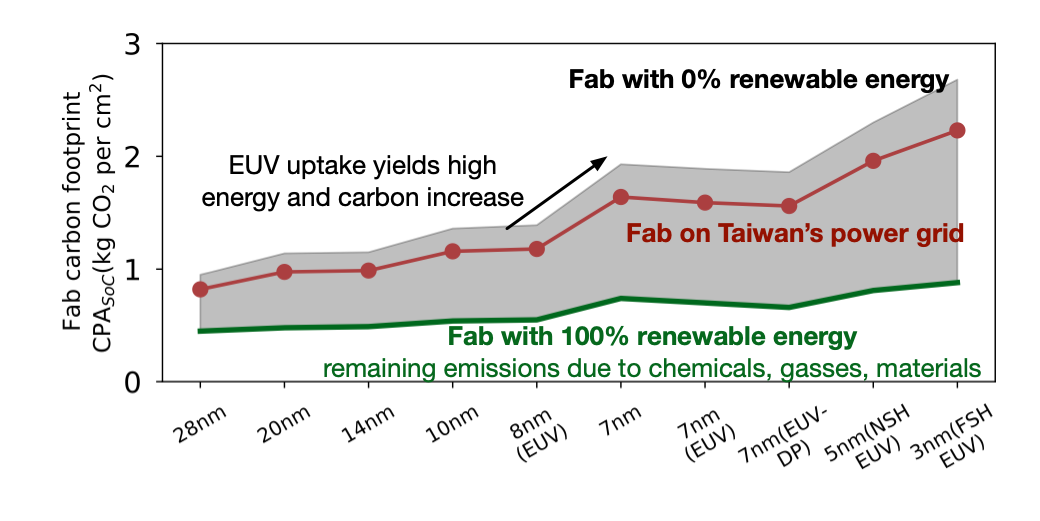}
\caption{Embodied carbon for semiconductor fabrication. Data from industry reports, device characterization \cite{act}.}
\label{fig:account-fab}
\end{minipage}
\hfill
\begin{minipage}[t]{0.48\linewidth}
\centering
\includegraphics[width=\textwidth,trim = 8mm 8mm 0mm 4mm, clip]{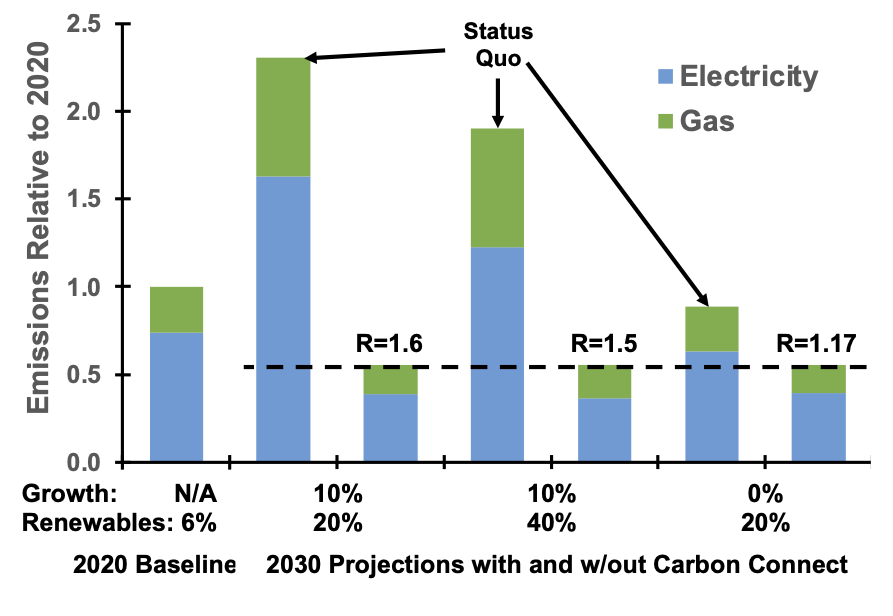}
\caption{Embodied carbon scenaris that vary fab electricity growth, renewable energy use, and 3R's of circular economy.}
\label{fig:embodied_future}
\end{minipage}
\end{figure}

Embodied carbon from semiconductor manufacturing is a major contributor to emissions~\cite{KLINE2019322, gupta21, PIRSON2021128966}, especially for mobile and embedded devices, due to high replacement rates and relatively low utilization. Nearly 75\% of Apple's corporate emissions are due to manufacturing~\cite{gupta21}. Billions of devices are expected to come online by 2027, and their embodied carbon may approach one gigaton of CO$_2$ per year, exceeding commercial aviation's footprint~\cite{PIRSON2021128966}. The largest semiconductor fabrication companies consume large amounts of electricity, especially for advanced technology nodes that require extreme ultraviolet lithography (Fig~\ref{fig:account-fab}). Their carbon costs increase even further when accounting for the gasses required by semiconductor manufacturing. 


Although fabs could reduce their emissions by using carbon-free energy from renewable sources, at present, carbon-free electricity is a meager 6\% of the total in Taiwan and South Korea, where most chips are produced. TSMC and Korea plan to increase their use of carbon-free energy to 40\% and 20\% of their respective totals by 2030~\cite{korea_renewables, tsmc_memory_website}. 

Fig-\ref{fig:embodied_future} presents several scenarios for embodied carbon, varying growth in fab energy demand and renewable energy supply. Even under optimistic assumptions in which fab demand is unchanged (0\%) and renewable energy supply increases by 20\%, the industry will miss its goal of reducing emissions by 45\%, which is illustrated by the dashed line. This outcome is partially explained by gases, which account for 25\% of total emissions and are unaffected by the use of renewable energy. Thus, reducing embodied carbon by 45\% requires more aggressive and innovative measures. 

Researchers will need to explore several mitigation strategies for embodied carbon, which arise from the three R's of circular economy---reduce, reuse, and recycle. Our analysis specifies an ``R-factor'' that estimates the extent to which these three R's are needed to reduce embodied carbon by 45\%. As an illustrative example, R=1.5 estimates the combined effect of reducing hardware procurement by 33\%, re-using hardware 1.5$\times$ longer, and recycling $1.5\times$ more hardware relative to 2020 levels. While different combinations are possible, parallel efforts to increase each of the three R's are essential to reaching the 45\% reduction target.


{\bf Reducing} hardware, architects should explore system design strategies that manufacture, provision, and allocate precisely the mix of hardware required for application needs. We need modular design tools for heterogeneous integration, the idea that hardware functions can be designed and implemented separately as small chiplets and then connected with fast interconnect networks \cite{coudrain19_active_interposr}. Chiplets are more carbon-efficient as fabs would manufacture precisely the required circuits and no more, reducing silicon area and improving manufacturing yield, which in turn reduces waste and carbon. Moreover, fabs could separate the manufacture of disparate capabilities---compute, memory, sensors---and use dedicated process flows for each, reducing the number of process steps and associated carbon. 

We also need datacenter-scale disaggregation, the idea that hardware could be organized into collections of network-attached components. Compute nodes would offer many CPUs but little DRAM, whereas memory nodes would offer the reverse. Disaggregation allows servers to independently scale a specific hardware type. "Lego-block" systems with custom core and memory configurations would better balance the system and amortize carbon, but designing such systems and then managing them at scale, where heterogeneity creates complexity, remains difficult. Such systems are more carbon-efficient than today's servers that inefficiently provision large quantitites of hardware in fixed proportions. For examples, today's servers provision many DRAMs for capacity but must also inefficiently provision a corresponding number of memory channels and processor sockets even when workloads underutilize the bundled bandwidth and compute \cite{malladi12}. 

{\bf Re-using} hardware in these decoupled systems, operators might replace components based on individual technology advances or failure rates rather than based on the fastest evolving or least reliable component, thereby extending the hardware's average tenure. Enabling component re-use improves sustainability by amortizing embodied carbon over a longer lifespan. Today, the typical server lifetime is three years, after which the entire rack is replaced with new hardware. Networking equipment lifetimes are longer, five years for switches/routers and ten years for the fiber cable plant, but periodic and wholesale replacement is still standard.

Disaggregation will benefit lifecycle management as separating the physical organization of resource types permits independent refresh and replacement. GPUs might refresh at a rate dictated by growing demands for AI workloads, whereas CPUs might refresh at a different rate tracking demand for general computation. Refresh based on individual technology advances or component failure rates rather than the fastest evolving or least reliable component will extend the datacenter's average component lifetime.

{\bf Recycling} hardware will require better instrumentation and health prediction to facilitate an efficient secondary market that disassembles systems into constituent components and sells them for a second life, further amortizing embodied carbon. Transparency is needed for market efficiency. For instance, heavily used processors from hyperscale datacenters will have very different resale values than lightly used processors from enterprise deployments. Thus, data must be curated by manufacturers, sellers, or third parties so that consumers can intelligently assign value to pre-owned hardware. We will design frameworks for registering hardware components and reporting their usage for statistical analysis. This would significantly expand the scope of economic activity for the semiconductor industry. Manufacturers might move toward leased equipment, which is known to be more highly utilized (and thus carbon-efficient) than owned equipment.

We draw inspiration from the role that odometers, vehicle history reports, and certified pre-owned designations play in the secondary vehicle market.  We envision hardware with ``odometers'' that account for their previous usage. The odometer will be implemented with dedicated, immutable, and tamper-resistant registers that count operations. A single odometer value is an imperfect proxy for history and additional measures are likely needed. For memories and disks, registers might count errors and faults in addition to read/writes. For all components, measures of physical conditions such as power variations, thermal stresses, and humidity will be helpful. Researchers will need to identify relevant features and develop compact representations, especially for longitudinal data such as temperature. 


\textbf{Risks.} One might argue that reducing, re-using, and recycling hardware runs counter to manufacturers' financial incentives. Manufacturers earn more revenue by selling more components, a major risk to sustainable design. Two factors mitigate this risk. Contrary to the historical commoditization of hardware components, modern datacenter operators are large customers that can demand and influence custom hardware and features, including those that enhance infrastructure efficiency and sustainability. Moreover, when successful, these research directions will produce a robust secondary market that expects reliability and provenance akin to the demand for cars with higher resale values and clear maintenance records. 

Others may question whether users would accept higher operational carbon from reused or recycled components. But these users could lower their life cycle carbon footprint and receive financial incentives like those found in ubiquitous, secondary markets for other capital equipment. Indeed, datacenter operators have tolerated less capable machines and greater operational complexity for financial reasons in the past (\textit{e.g.} Google's cluster with commodity servers \cite{barroso03google}) and may do so for carbon in the future. 

\section{Operational Carbon}

\begin{figure}
\begin{minipage}[t]{0.48\linewidth}
\centering 
\includegraphics[width=\textwidth, trim=0mm 3mm 0mm 5mm, clip]{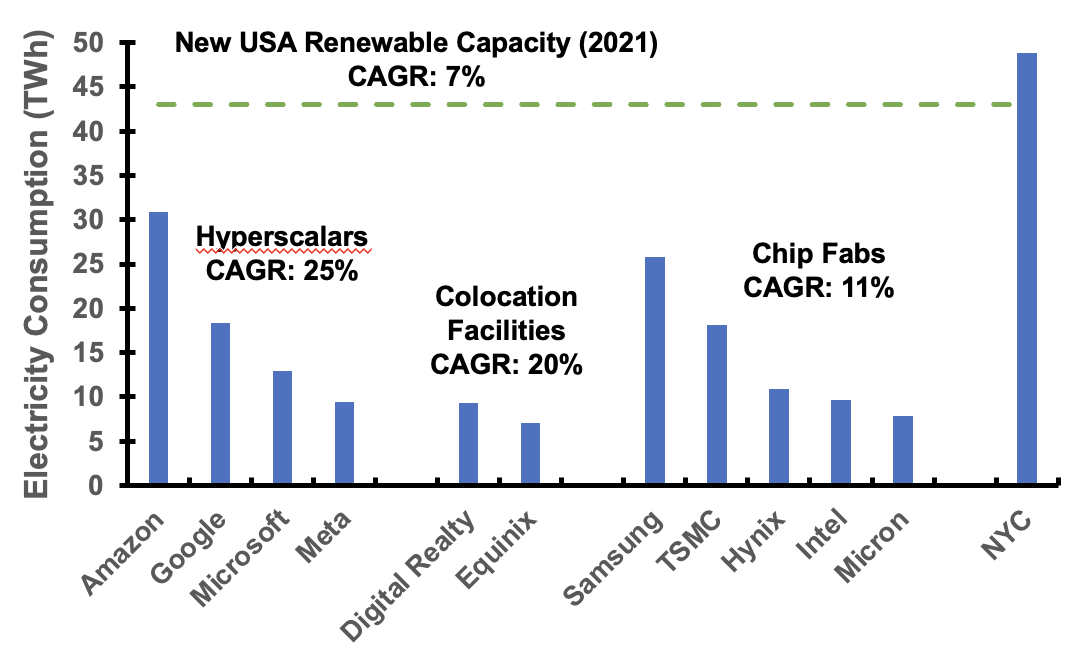}
\caption{Electricity usage (2021) for datacenter and fabrication facilities. CAGR growth: 2015 to 2021. Corporate sustainability reports, EIA, and \cite{gupta21}.}
\label{fig:account-ict}
\end{minipage}
\hfill
\begin{minipage}[t]{0.48\linewidth}
\centering
\includegraphics[width=\textwidth,trim = 3.2mm 0mm 0mm 4mm, clip]{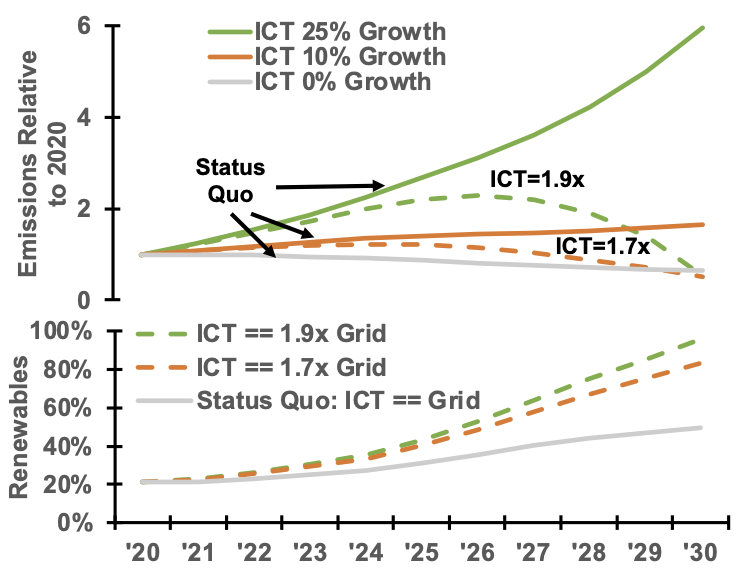}
\caption{Operational carbon reduction (45\% by 2030) achieved via 1.7x higher uptake in ICT renewable electricity compared to the grid average.}
\label{fig:operational_future}
\end{minipage}
\end{figure}

Over the next decade, annual ICT energy demand is projected to exceed 100 exajoules, reaching nearly 15\% of the world's energy production~\cite{src_decadal_plan}. This explosive growth is driven by diverse applications such as artificial intelligence, virtual spaces, Internet of Things, blockchains, \textit{etc}. Electricity use at Google, Meta, and Microsoft grew at 25\% per year from 2015 to 2021, nearly quadrupling over this period. In contrast, U.S. investments in renewable energy grew at only 7\% per year (Fig-\ref{fig:account-ict}).  In 2021, hyperscale datacenters consumed an additional 19 TWh compared to 2020---nearly half of the 44 TWh of new renewable capacity that came online that year.  

Our analysis of operational carbon highlights the essential role of renewable energy for computing (Fig-\ref{fig:operational_future}). Suppose renewable energy capacity grows at 10\% per year as forecasted by the U.S. Energy Information Adminstration. In a conservative scenario, computing's energy demand remains constant at 2020 levels and renewable energy's growth would reduce carbon by 36\%. However, in more realistic scenarios that reflect industry consensus \cite{src_decadal_plan}, computing's energy demand increases by 10\% to 25\% per year and renewable energy's growth struggles to keep pace. Meeting these demand yet reducing carbon by 45\% requires computing to adopt renewable energy 1.7--1.9$\times$ faster than the U.S. average. 

\textbf{Demand Response.} If renewable energy accounts for a large majority of the total by the middle of the century, intra-day supply variability will require changes to when and where computation is performed. Sustainable datacenters will need to delay and boost computation when carbon-free energy is scarce and abundant, respectively, while still meeting strict peformance requirements. Such schedules would require re-thinking conventional wisdom in which datacenters compute constantly at peak power to amortize costs of building the facility and provisioning its power \cite{fan07}.



Datacenters will need sophisticated demand response (DR) frameworks that modulate energy demand in response to carbon-free energy supply. DR will require coordination between the energy grid, datacenter operator, and datacenter users, and it will require hardware and software mechanisms that trade-off performance and power. Ideally, DR frameworks will both incentivize participation and guarantee service. Game theory might provide rigorous foundations for modeling and shaping system dynamics when users independently and selfishly pursue performance goals. Real-time scheduling and robust machine learning  might ensure decisions satisfy diverse service obligations.

\textbf{Abstractions and Interfaces.}
We envision abstractions between the energy grid, datacenter operator, and datacenter users. Operators will interface with the grid, analyzing the dispatch stack, energy prices, and carbon intensity to set datacenter power budgets on an hourly basis based on sustainability objectives. Users and their jobs interface with the datacenter to receive power allocations without being exposed to the grid's complexity. Each user must define their own mechanisms for modulating power and computing within its allocation. 

The datacenter will receive information about energy supply through concise interfaces. One interface would communicate real-time prices that incentivize datacenters to modulate energy use. This scenario would efficiently match supply and demand, but departs from today's contracts that charge based on the amount of provisioned power rather than actual use. And what prices are required to achieve the desired demand response? An alternative interface communicates carbon intensity rather than price. But this scenario assumes datacenters would modulate demand to reduce operational carbon without compensation. 


\textbf{Power Modulation.}
Each user must define and implement multiple operating modes that modulate power when required. Hardware mechanisms will rely on energy proportionality, the idea that power should rise and fall with workload. Energy-proportional hardware is difficult to design because most components have a significant fixed power cost dissipated even at near-zero load. Decades of research have improved CPUs, but today's datacenters deploy large memory systems and graphics processing units that will need to be designed for energy proportionality. Memory will need new interfaces as today's dissipate high fixed power to deliver high bandwidth. Accelerators will need better support for virtualization and resource sharing, which better amortizes fixed power costs over more useful computation. 

Software mechanisms will rely on approximate, degraded computing. Online applications implement contingency plans for site events, ensuring varying degrees of service that depend on system availability and downtime. We will explore real-time system design and anytime algorithms to provide a smoother spectrum of trade-offs between quality and power than permitted in today's systems. This approach generalizes the search engine's strategy of delivering the most relevant results within some allotted time \cite{reddi10}. Strategies for computational sprinting allow workloads to dynamically consume additional resources as power budgets permit \cite{fan16}.


\textbf{Intelligent Decisions.} A cognitive stack would permit the separation of concerns and clean abstractions. The stack organizes power management into a fast, low-level reactive layer that is vertically coupled to a strategic, high-level deliberative layer. An agent monitors local job performance and hardware utilization, optimizing its power requests to achieve its performance goals while accounting for global datacenter conditions and competition from other agents. The reactive policy could adjust a processor’s power mode in response to program phases while the deliberative policy ensures each processor’s adjustment anticipates other processors’ policies and the datacenter’s broader goals in sustainability, safety, and stability. 

The cognitive stack could leverage multi-agent game theory and reinforcement learning for dynamic decision making \cite{yeh24}. Dynamism is important because computation exhibits time-varying behavior, and allocation decisions in the present should account for the past and anticipate the future. For example, consider a repeated game in which agents spend tokens for power. Each agent learns a policy for spending tokens, requesting power, and mapping power to datacenter resources. When carbon-free energy is scarce, the datacenter could offer more tokens to users that defer jobs or require more tokens from those who compute. How should agents spend tokens to maximize long-term performance when allocations in one time period affect those in an uncertain future? How should the datacenter price power to achieve sustainability or DR goals?



\textbf{Risks.} Some might wonder whether DR will be necessary given how datacenter operators are investing in renewable energy. Today, net zero claims rely on wind/solar energy investments that offset datacenter energy use, but carbon-free energy is too often generated at times and places that do not align with when and where computation happens such that "net-zero" datacenters consume carbon-intensive energy in many hours of the year \cite{acun23}. Others might wonder whether DR already exists. Today's narrow solutions focus on stability as grids request reduced use to avoid rare power emergencies, but even these suffer from incentive problems \cite{wierman14}. Researchers have studied how datacenters could participate in DR using simplifying assumptions. For example, studies assume 20\% of power is consumed by batch jobs that are deferrable within a 24-hour window without loss of utility \cite{wierman14, radovanovic21}, but practical DR must accommodate diverse mixes of heterogeneous jobs. 

\section{Energy Economics}

Economics and policy shape the solution space for carbon-efficient computing. Governments might implement carbon trading or incentives for low-carbon energy. The private sector might implement offset programs, leading to renewable energy purchase agreements and credits. Future demand response frameworks will require sophisticated markets that price electricity at its true social marginal cost and incentivize users to schedule computation accordingly. Although there is extensive literature on low-carbon policies for other industries \cite{abrell19}, economic analysis of policies specifically aimed at computing is relatively unexplored. In the near term, industry will benefit from policy-induced incentives when investing in renewable energy supply (e.g., renewable energy certificates). But as supply grows, industry must be able to monetize its flexibility in energy demand. Datacenters are often the largest consumers on the grid and we must understand how their locally optimized decisions for net zero operations will affect other consumers and impact society. 

Furthermore, we will study how improved sustainability impact demand for computing. Given an unpriced environmental externality \cite{hardin68}, such as carbon, one might ask whether society is computing too much. What is the optimal amount of computing for society? Will more efficient algorithms and systems lead to so much demand for new computing applications that overall carbon will increase? The Jevons Paradox states increased efficiency may not reduce demand for energy in the long run (and may even increase it). Prior research suggests, as a technology becomes more efficient, its use increases and produces rebound effects that range from 10\% to 40\%, reducing but not eliminating energy savings \cite{gillingham16}. But there has been no study of these effects for computing. 

We would need to estimate three types of rebound effects as technological efficiency lowers operating costs. First, direct effects arise when lower costs increase use of the technology. Datacenters likely exhibit strong direct effects as operators provision hardware to maximally use provisioned power \cite{fan07}; more efficient processors lead to datacenters with more processors.  Second, indirect effects arise when lower costs increase use of other technologies. Quantifying indirect effects requires understanding substitutability and complementary between hardware components, which in turn depend on hardware capacity translates into software performance; more efficient processors may lead to datacenters with more memory as well. Finally, macroeconomic effects arise when lower costs encourage technology use for new applications. Efficient processors may scale the use of large AI models for everyday tasks (\textit{e.g.}, conversational bots) rather than niche tasks (\textit{e.g.}, playing games).

\section{Conclusion}
\label{sec:conc}

Addressing the sustainability challenge requires a broad community. By redefining the way researchers in computing consider environmental sustainability, researchers will establish new standards for carbon accounting in the computing industry, thereby influencing future energy policy and legislation. An interdisciplinary community of researchers dedicated to sustainable computing is needed to train the next generation of innovators in the combined fields of computer science, electrical engineering, industrial ecology, and energy policy. Academic-industry partnerships are needed to accelerate the adoption of sustainable computing practices.  

The research community must seek coordinated solutions to reduce the carbon footprint of information and communication technology by 45\% within the next decade. These solutions must include methods transparent, accurate carbon accounting. They must include strategies for carbon-efficient system design, intelligent power management, and hardware life cycle management. And they must lead to infrastructure that supports rapidly growing capabilities and applications such as artificial intelligence. A shift towards sustainability could spark a transformation in how computer systems are manufactured, allocated, and consumed, thereby establishing foundations for a future of continued advances in high-performance, sustainable computing.

\bibliography{biblio}
\bibliographystyle{abbrv}

\end{document}